\documentclass[conference]{IEEEtran}
\IEEEoverridecommandlockouts
\usepackage{cite}
\usepackage{amsmath,amssymb,amsfonts}
\usepackage{amsthm}
\usepackage{breqn}
\usepackage{algorithmic}
\usepackage{graphicx}
\usepackage{textcomp}
\usepackage{xcolor}
\usepackage[hidelinks]{hyperref}
\def\BibTeX{{\rm B\kern-.05em{\sc i\kern-.025em b}\kern-.08em
		T\kern-.1667em\lower.7ex\hbox{E}\kern-.125emX}}

\newtheorem{proposition}{Proposition}
\usepackage{xspace}

\makeatletter
\DeclareRobustCommand\onedot{\futurelet\@let@token\@onedot}
\def\@onedot{\ifx\@let@token.\else.\null\fi\xspace}

\def\ie{\emph{i.e}\onedot}

\makeatother

\DeclareMathOperator{\Tr}{Tr}

\DeclareMathOperator*{\argmin}{arg\,min}

\begin{document}
	\title{Denoising bivariate signals \\ via smoothing and polarization priors
		\thanks{The emails of the authors are 
		\{firstname.lastname\}@univ-lille.fr: \{yusuf-yigit.pilavci, jeremie.boulanger, pierre-antoine.thouvenin, pierre.chainais\}}
	}
	\author{\IEEEauthorblockN{Yusuf Yi\u{g}it P\.{I}LAVCI, J\'er\'emie BOULANGER, Pierre-Antoine THOUVENIN, Pierre CHAINAIS} 
		\IEEEauthorblockA{\textit{Univ. Lille, CNRS, Centrale Lille, UMR 9189 CRIStAL, F-59000}, Lille, France}
	}
	\maketitle
	
	\begin{abstract}
		We propose two formulations to leverage the geometric properties of bivariate signals for dealing with the denoising problem. 
		In doing so, we use the instantaneous Stokes parameters to incorporate the polarization state of the signal. 
		While the first formulation exploits the statistics of the Stokes representation in a Bayesian setting, the second uses a kernel regression formulation to impose locally smooth time-varying polarization properties. 
		In turn, we obtain two formulations that allow us to use both signal and polarization domain regularization for denoising a bivariate signal. 
		The solutions to them exploit the polarization information efficiently as demonstrated in the numerical simulations.   
	\end{abstract}
	
	
	\section{Introduction}
	The inverse problems involving bivariate signals,~\ie~time series with two components, are at the very heart of various phenomena in nature. 
	From cosmology to underwater acoustics, seismology to optics or neuroscience to oceanography, one needs to deal with bivariate signals which require the joint analysis of the components of the data~\cite{flamant2018approche}.  
	In doing so, the bivariate signals have a particular place in the realm of multivariate signal processing where one can find several representations to interpret the joint relationships of the two components. 
	On one hand, one has the covariance matrix which is the classical tool for any multivariate signal.   
	On the other, one may choose the Stokes representation, which is inspired by the optics~\cite{stokes1851composition}. 
	While these two representations are equivalent and connected through a set of non-linear equations, in the latter, one can interpret the joint variation of the two components as the \emph{polarization state} of a bivariate signal. 
	In addition, the geometric properties of the bivariate signals are much simpler to interpret from the Stokes representations. 
	\smallskip
	
	\begin{figure}
		\includegraphics[width=0.48\linewidth,trim={2cm 0cm 2.5cm 2cm},clip]{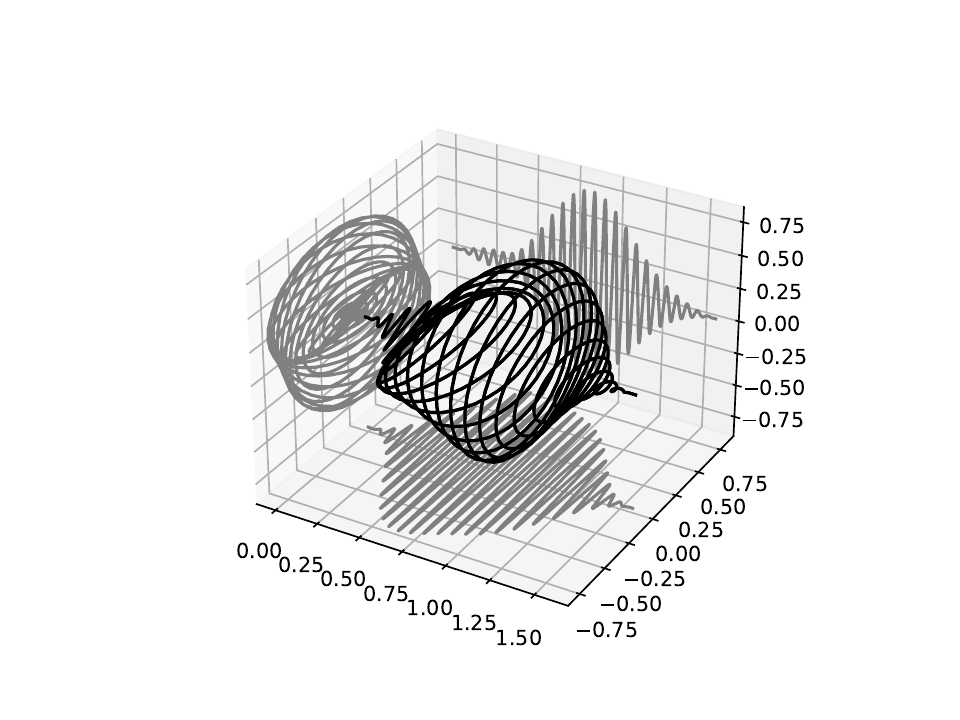}
		\includegraphics[width=0.48\linewidth,trim={2cm 0cm 2.5cm 2cm},clip]{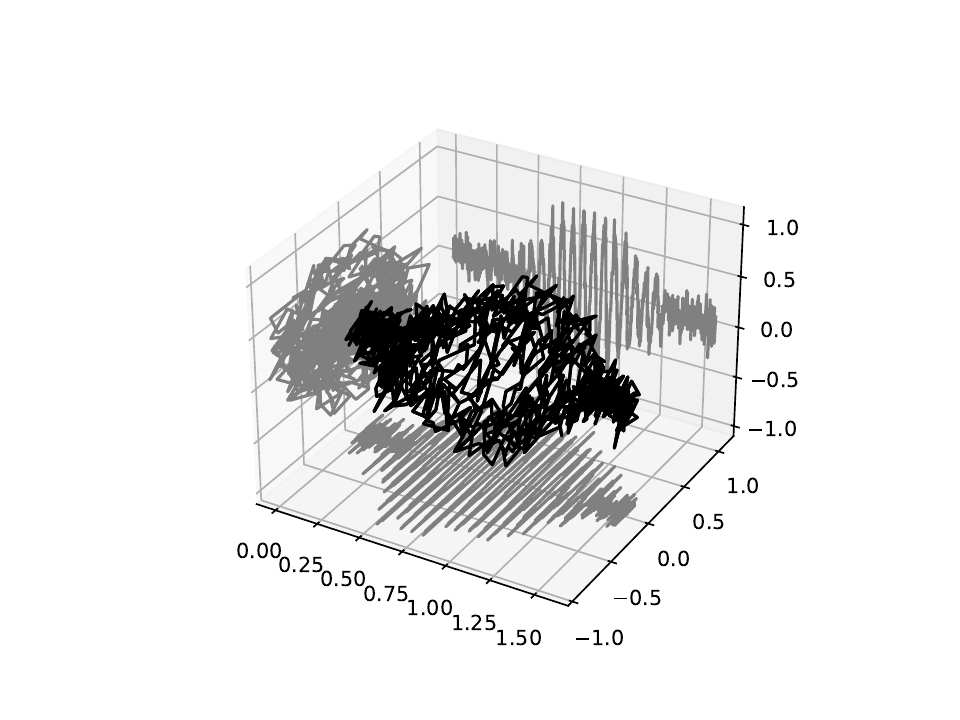}
		\caption{An example of a polarized bivariate signal (at left) and its contaminated version by noise (at right). The SNR of the noisy signal is $9.75$. }
		\label{fig:bivariate-example}
	\end{figure}
	
	\noindent \textbf{Related works.} 
	A notable example of bivariate signals in astrophysics is the analysis of gravitational waves~\cite{cano2022outils}. 
	Several problems in their analysis such as their characterization~\cite{cano2022outils}, generation~\cite{cano2022modeles} or reconstruction~\cite{cano2023contrainte} take the form of an inverse problem where one needs to jointly analyze two signal components and requires the proper tools. 
	Especially in~\cite{cano2023contrainte}, one considers the reconstruction of gravitational waves by constraining the polarization.  
	Their assumption is that the noisy part of the signal is not polarized unlike the true signal behaves as a polarized AM-FM and thus has a particular polarization state, which means that the joint variation of the signal follows a particular pattern.
	The main limitation comes from the fact that one needs to know the polarization of the true signal in advance.   
	
	This paper deals with solving a particular inverse problem, which is signal denoising (See Fig.~\ref{fig:bivariate-example}).
	Although it is one of the most basic inverse problems, the adaptation of the Stokes representations within the classical denoising formulations has not been studied in detail.  
	The question of how to integrate the polarization state of the signal for improving the performance (the denoising performance in this case) is of central importance.   
	As an attempt to address this question, we propose two formulations for bivariate signal denoising where the instantaneous Stokes parameters are used. 
	The first framework formulates the problem at hand in the Bayesian setting in which one needs to choose the proper likelihood and priors for the individual components of the bivariate signal and its Stokes representations. 
	The second approach deploys the kernel regression method to use the normalized Stokes parameters.
	In both cases, the resulting optimization problem turns out to be differentiable. 
	
	Section~\ref{sec:prob} gives the problem definition along with the underlying noise assumption. 
	Then, section~\ref{sec:background} provides the necessary background on the instantaneous Stokes parameters. 
	Section~\ref{sec:method} details the proposed frameworks to formulate and solve the problem at hand. In Section~\ref{sec:experiments}, we illustrate these methods in numerical simulations and discuss the results\footnote{The codes to reproduce the figures of this paper can be found in~\url{{https://gitlab.cristal.univ-lille.fr/ypilavci/denoising-polarization}}}. Finally, we conclude in Section~\ref{sec:conc}.  
	\begin{figure}
		\centering
		\includegraphics[width=\linewidth]{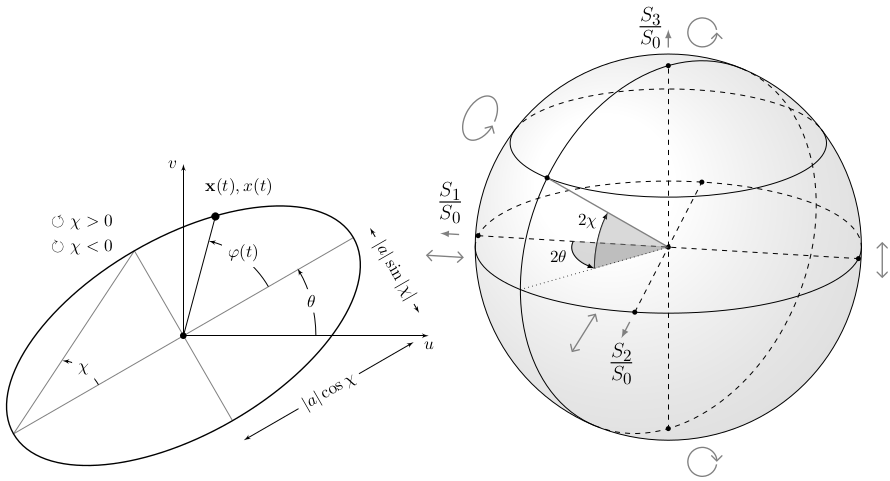}
		\caption{The polarization ellipse~\ie~the ellipse drawn by the bivariate signal in $u$-$v$ axis (at left) and the Poincar\'e sphere (at right). Taken from~\cite{flamant2018approche}.}
		\label{fig:poincare}
	\end{figure}
	
	\section{Problem definition}
	\label{sec:prob}
	Given a bivariate signal $x(t)$ with two real-valued components $u(t)$ and $v(t)$, the vector representation is $x(t)=[u(t), v(t)]^\top\in\mathbb{R}^{2}$.
	In the context of the denoising problem, one aims to recover the (discrete) bivariate signal $\mathbf{x}=[{x}(t_1),\dots,{x}(t_N)]^\top$ from given noisy measurements $\mathbf{y}=[y(t_1),\dots,y(t_N)]^\top$ at the timestamps $t_1,\dots,t_N$. 
	The measurement model is: 
	\begin{equation}
		\mathbf{y} = \mathbf{x} + \boldsymbol{\epsilon}, 
	\end{equation}
	where $\epsilon(t_i) = [n_{u}(t_i), n_v(t_i)]^\top\sim\mathcal{N}(\mathbf{0},\sigma^2\mathsf{I}_{2\times2})$. 
	Typically, one uses the following generic Bayesian framework to tackle these types of problems: 
	\begin{equation}
		\mathbf{x}^{\star} = \argmin_{\mathbf{x}\in\mathbb{R}^{2\times N}} f(\mathbf{x},\mathbf{y})+  g(\mathbf{x})
	\end{equation}
	where $f$ is the data fidelity loss function and $g$ is the regularization penalty. 
	The design of $f$ and $g$ is closely related to the underlying likelihood and prior distributions, respectively. 
	However, this type of formulation alone does not necessarily regulate the cross-component properties which is often informative in bivariate signal processing. 
	This paper presents two ways to extend this formulation in order to leverage the polarization properties. 
	To do so, we first go through the necessary background on the instantaneous Stokes parameters in the next section. 
	%	The Bayesian posterior distribution encodes our knowledge on the true signal $\mathbf{x}$, by combining the likelihood and the prior distributions. 
	%	Given the likelihood $\mathbb{P}(\mathbf{y}|\mathbf{x})$ and the prior $\mathbb{P}(\mathbf{x})$ probability distributions, the purpose is to find the estimate $\mathbf{x}$ that maximizes the posterior distribution $\mathbb{P}(\mathbf{x}|\mathbf{y})$. By using the Bayes' rule, one can reformulate the optimization problem as follows: 
	%	Given the additive Gaussian noise model, the corresponding likelihood function $\mathbb{P}(\mathbf{y}|\mathbf{x},\sigma)$ is indeed the Gaussian probability density function with the mean $\mathbf{x}$ and the variance $\sigma^2$. This leads us to $\|x-y\|_2^2$ as the fidelity term. Then, one only needs to decide on the prior distribution $\mathbb{P}(\mathbf{x})$ that encodes the structural information of the underlying signal. 
	
	\section{Instantaneous Stokes parameters}
	\label{sec:background}
	To define the instantaneous Stokes parameters, we first define the bivariate equivalent of the complex analytic signal of $x(t)$~\ie~the complex signal obtained by removing negative frequencies from the original signal. 
	By analogy, the complex analytic signal of $x(t)$ reads: 
	\begin{equation}
		x_{a}(t) = \begin{bmatrix}
			u(t) +\boldsymbol{i}\hat{u}(t) \\ 
			v(t) +\boldsymbol{i}\hat{v}(t)
		\end{bmatrix} \in \mathbb{C}
	\end{equation}
	where $\hat{u}(t)$ and $\hat{v}(t)$ are the Hilbert transform of $u(t)$ and $v(t)$. The analytic signal $x_a(t)$ has the following connection with the instantaneous Stokes parameters:
	\begin{equation}
		x_{a}(t)\overline{x_{a}(t)} = \begin{bmatrix}
			S_0(t) + S_1(t) & S_2(t) + \boldsymbol{i}S_3(t) \\ 
			S_2(t) - \boldsymbol{i}S_3(t)& 	S_0(t) - S_1(t)  \\
		\end{bmatrix},
	\end{equation}	
	where $\overline{x_{a}(t)}$ is the complex conjugate of ${x_{a}(t)}$.
	The instantaneous Stokes parameters are completely energetic quantities and $S_0(t)$ is the instantaneous intensity while $S_{1,2,3}(t)$ relate to the time-varying geometric properties under narrowband assumptions as one has: 
	\begin{equation}
		\begin{split}
			S_0(t) &= a(t)^2, \\ 		
			S_1(t) &= a(t)\cos(2\chi(t))\cos(\theta(t)), \\ 
			S_2(t) &= a(t)\cos(2\chi(t))\sin(\theta(t)), \\ 
			S_3(t) &= a(t)\sin(2\chi(t)).
		\end{split}
	\end{equation}
	where $a(t)$ is the instantaneous amplitude of the signal. 
	The angles $\chi(t)$ and $\theta(t)$ are the angles of the ellipse drawn by the signal in $u$-$v$ axis (See Fig.~\ref{fig:poincare}).
	Note that the Stokes parameters are independent from the instantaneous phase $\phi(t)$ therefore one cannot rely only upon the Stokes parameters for a complete representation of a bivariate signal.
	The connection of the ellipse angles with the Stokes parameters is even more prominent in the normalized Stokes parameters which read: 
	\begin{equation}
		\mathbf{s}(t)=[s_1(t),s_2(t),s_3(t)]^\top = \left[\frac{S_1(t)}{S_0(t)},\frac{S_2(t)}{S_0(t)},\frac{S_3(t)}{S_0(t)}\right]^\top 
	\end{equation}
	and for deterministic signals, one always verifies that ${S_1(t)}^2+{S_2(t)}^2+{S_3(t)}^2 = {S_0(t)^2}$. As a result, these parameters can be considered as coordinates that live on the sphere and the spherical coordinates correspond to the ellipse angles as shown in Fig.~\ref{fig:poincare}. In optics, this sphere is often called Poincar\'e sphere and the spherical coordinates have a direct interpretation in terms of the ellipse parameters of the bivariate signal~\cite{flamant2018approche}.

	%	For example, choosing $\mathbb{P}(\mathbf{x})$ as $\mathcal{N}(0,1/\lambda)$ or $\mathcal{L}(0,1/\lambda)$ (Laplace or doubly-exponential distribution) leads to adding the ridge $\lambda\|\mathbf{x}\|_2$ or the Lasso (sparsity)  $\lambda\|\mathbf{x}\|_1$ regularization term, respectively.
	%	In this paper, our purpose is to show how to leverage the Bayesian modelling in order to incorporate the Stokes representation of the bivariate signals.  
	\section{Denoising with the Stokes parameters}
	\label{sec:method}
	
	\subsection{Bayesian formulation with the Stokes parameters}  
\begin{figure}
		\hspace{1cm}
		\includegraphics[width=0.7\linewidth]{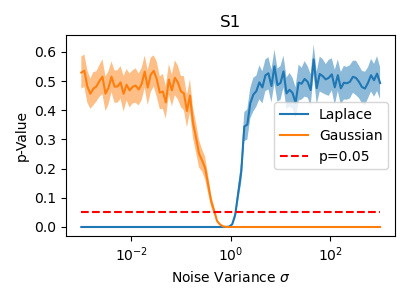}
		\caption{The comparison of the empirical distribution of the geometrical Stokes parameters with the theoretical distributions for varying noise variance. Only the comparison for $S_1$ is plotted however the results for $S_2$ and $S_3$ are  By using the Kolmogorov-Smirnov test, we measure the fitness of the theoretical distributions of Gaussian and Laplace distribution with to the empirical histograms. The fitness is given in $p$-value. A higher $p$-value indicates better fitness.}
		\label{fig:sigma-vs-pvalue}
	\end{figure}

	Let us define the discrete instantaneous Stokes parameters as $\mathbf{S}^{x}_{j}=[{S}^{x}_{j}(t_1),\dots,{S}^{x}_{j}(t_N)]$, $j=0,\dots,3$, for the signal $x(t)$ and the timestamps $t_1,\dots,t_N$. 
	In order to formulate the problem with the Stokes parameters, one needs to derive the joint likelihood distribution $\mathbb{P}(\mathbf{S}^{y}|\mathbf{S}^{x})$ which is not directly tractable. A naive but tractable approximation is to analyze the likelihood per Stokes parameter separately. 
	Interestingly, these distributions take different forms for the intensity $S_{0}$ and the geometry-related Stokes parameters $S_{1,2,3}$. 
	In our derivations, the key property that we use is given by the following proposition. 
	\begin{proposition}
		Let $n(t)\sim\mathcal{N}(m(t),\sigma)$ be a Gaussian process and $\hat{n}(t)$ be its Hilbert transform. Then, $n(t)$ and $\hat{n}(t)$ are independent Gaussian processes.
		\begin{proof}
			See Chapter 6 of~\cite{picinbono1993random}. 
		\end{proof} 
	\end{proposition}
	\smallskip
	\noindent \textbf{Likelihood.} Then, let us examine the random components of the Stokes parameters one by one by using this proposition. Firstly, the intensity-related Stokes parameter reads: 
	\begin{equation}
		S_0^y(t) = U(t)^2 + V(t)^2 + \hat{U}(t)^2 + \hat{V}(t)^2, 
	\end{equation}
	where $U(t)$, $V(t)$, $\hat{U}(t)$ and $\hat{V}(t)$ are independent Gaussian random processes due to Prop. 1 with the means $u(t)$, $v(t)$, $\hat{u}(t)$ and $\hat{v}(t)$ and $\sigma^2$ variance. As a result, $S_0^y(t)\sim\sigma^2\chi^2_4\left(\frac{S_0^x(t)}{\sigma^2}\right)$ is a scaled random process with $\chi^2$ distribution with the non-centrality parameter $\frac{S_0^x(t)}{\sigma^2}$ and the degree parameter $4$. Then, the negative log-likelihood function $-\log\mathbb{P}({S}^{y}_{0}(t)|{S}^{x}_{0}(t)=-\ell(y|x)$ reads: 
	\begin{equation}
		\ell(y|x) = \frac{{S}_0^x(t)}{2\sigma^2} + \frac{1}{2}{\log{S}_0^x(t)} - \log I_1\left(\frac{\sqrt{S_0^x(t)S_0^y(t)}}{\sigma^2}\right),
		\label{eq:stokes0_stat}
	\end{equation}
	where $I_1$ is the first-kind Bessel function.
	A very similar analysis can be done for the geometry-related Stokes parameters: 
	\begin{equation}
		\begin{split}
			S_1^y(t) &= U(t)^2 - V(t)^2 + \hat{U}(t)^2 - \hat{V}(t)^2, \\ 
			S_2^y(t) &= 2(U(t)V(t) - \hat{U}(t)\hat{V}(t)), \\ 			
			S_3^y(t) &= 2(\hat{U}(t){V}(t) - {U}(t)\hat{V}(t)), \\			
		\end{split}
	\end{equation}
	After several algebraic manipulations, each Stokes parameter turns out to have the same distribution which we denote by $S_{i}^{y}(t)\sim\mathcal{M}_{i}(S_{i}^x(t),S_{0}^x(t),\sigma)$. 
	A closer look at the components of this distribution shows that for $i=1,2,3$: 
	\[
	S_{i}^{y}(t) = S_{i}^{x}(t) + N_i(t) + L_i(t) \sim\mathcal{M}_{i}(S_{i}^x(t),S_{0}^x(t),\sigma), 
	\]
	where $N_i(t)\sim\mathcal{N}(0,4\sigma^2 S_{0}^x(t))$ and $L_i(t)\sim\mathcal{L}(0,2\sigma^2 )$ (Laplace distribution) are correlated. Moreover, for high values of $\sigma^2$ with respect to the signal intensity $S^x_0(t)$, $L_i(t)$ becomes the dominant component whereas for smaller values of $\sigma$, $N_i(t)$ is the dominant component (See Fig.~\ref{fig:sigma-vs-pvalue}).  
	As a result, the corresponding log-likelihood function does not come in an exact closed form. 
	However, depending on the assumption of high or low SNR of the measurements, the log-likelihood function can be better approximated by that of the Laplace distribution,~\ie $\|\mathbf{S}^x_i-\mathbf{S}^y_i\|_1$ or that of the Gaussian distribution,~\ie $\|\mathbf{S}^x_i-\mathbf{S}^y_i\|_2^2$ for $i=1,2,3$, respectively. In the rest, we use the latter assuming that the signal is not dominated by the noise level.    
	\smallskip
	
	\noindent \textbf{Priors.} We assume that both the true signal and the Stokes parameters vary smoothly through time. 
	A typical way to regularize the variation over time is by the penalty term $\lambda_1\mathbf{x}^\top\mathsf{L}\mathbf{x}$ where $\mathsf{L}$ is the discrete Laplacian operator. 
	Adapting this penalty term for the Stokes parameters yields $\lambda_s(\mathbf{S}^x_{n})^\top\mathsf{L}\mathbf{S}^x_{n}$ for $n=0,\dots,3$. 
	Similarly, this penalty term encodes our assumption of the smooth evolution of the polarization properties over time. 
	In addition, these terms can be seen as assuming Gaussian prior distributions on the finite distances $x(t_{i+1})-x(t_{i})$ or ${S}^x_{n}(t_{i+1})-{S}^x_{n}(t_{i})$ in the Bayesian setting.  
	In the rest, we use these regularization penalty terms. 
	\smallskip
	
	\noindent \textbf{Mixed MAP estimator.} 
	As a result, the loss function $f_1(\mathbf{x},\mathbf{y})$ derived from the posterior distribution in the signal domain reads: 
	\begin{equation}
		-\log\mathbb{P}(\mathbf{x}|\mathbf{y}) \propto  f_1(\mathbf{x},\mathbf{y})=\|\mathbf{x} - \mathbf{y}\|^2_2  + \lambda_1\Tr(\mathbf{x}^\top\mathsf{L}\mathbf{x}). 
	\end{equation}
	Similarly, in the Stokes domain, the corresponding loss function $f_2(\mathbf{x},\mathbf{y})$ reads: 
	\begin{dmath}
		f_2(\mathbf{x},\mathbf{y})= \beta_1\sum_{k=1}^{N}\ell(y(t_k)|x(t_k)) + \beta_2\sum_{i=1}^3\|\mathbf{S}_{i}^x-\mathbf{S}_{i}^y\|^2_2  + \lambda_s\sum_{i=0}^3(\mathbf{S}_{i}^x)^\top\mathsf{L}\mathbf{S}_{i}^x,
	\end{dmath}
	where $\ell(y(t_k)|x(t_k))$ is the negative log-likelihood function for $S_0^y(t)|S_0^x(t)$ defined in Eq.~\eqref{eq:stokes0_stat}, $\beta_{1}$, $\beta_2$, $\lambda_1$ and $\lambda_s$ are the hyperparameters to regulate the trade-off between the fidelity and regularization terms. 
	Then, the mixed problem formulation is
	\begin{equation}
		\mathbf{x}^{\star}_{mix} = \argmin_{\mathbf{x}\in\mathbb{R}^{2\times N}} f_1(\mathbf{x},\mathbf{y}) + 	f_2(\mathbf{x},\mathbf{y})
		\label{eq:firstprob}
	\end{equation}  
	The resulting loss function is differentiable thus any gradient descent algorithm is applicable. 
%	The non-convexity mainly comes from the fact that one needs to optimize the two components of $\mathbf{x}$ for the loss function that includes quartic terms of $\mathbf{x}$ due to the involvement of the Stokes parameters. 
	%	Nevertheless, we keep in mind that more sophisticated algorithms to tackle this problem can be developed by combining more adapted tools such as quaternion optimization~\cite{flamant2021general} which we leave for future work for now.   
	
	%	Given this formulation, indeed the main question is if we gain or lose anything in terms of the final performance. Intuitively, one can directly see that the signal and the Stokes domain allow us to analyze the bivariate signal in different axes. 
	%	By imposing smoothness prior to the signal domain, the variations in the two components are individually regulated. 
	%	On the other hand, the Stokes representation allows us to regulate the variations of two components w.r.t. each other that encodes the polarization properties of the signal. 
	%	Therefore, incorporating these two axes, one would expect to obtain an improved performance in denoising (or any inverse problem). The numerical examples in Section~\ref{sec:experiments} also confirm this expectation.  	

	\subsection{Denoising by smoothing the normalized Stokes parameters}
	
%	Although, both the unnormalized and normalized Stokes parameters are equivalent representations, a natural question here is whether the normalized Stokes along with the intensity parameter $[S_0(t),s_1(t),s_2(t),s_3(t)]$ is more or less useful in solving of the inverse problem at hand. 
%	As an attempt to answer this question, we seek an alternative formulation of the denoising problem both in the signal $x(t)$ and the normalized Stokes domain $[S^x_0(t),s^x_1(t),s^x_2(t),s^x_3(t)]$ while we take the spherical nature of the normalized parameters into account. 
	
	The likelihood function for the vector of the normalized Stokes parameters $\mathbf{s}^x$ is not easily tractable and does not come in a closed form. 
	Therefore, we propose to change our Stokes-related loss function to the following approximation by getting inspired by the kernel regression methods~\cite{hastie2009elements}: 
	\begin{equation}
		\begin{split}
			\ell_k(\mathbf{x}|\mathbf{y},\gamma) &= \sum_{i=1}^N w(i)\sum_{j=i}^{i+W} K_\gamma(t_i,t_j)d(\mathbf{s}^x(t_i),\mathbf{s}^y(t_j)), \\ 
			\text{with}&\quad w(i)=	\sum_{j=i}^{i+W} K_\gamma(t_i,t_j)S_0^x(t_j),		
		\end{split}
	\end{equation} 
	where $W<<N$ is the size of local window,  $K:\mathbb{R}^2\rightarrow\mathbb{R}$ is the kernel function, usually the Gaussian kernel $K_\gamma(t_i,t_j)= \exp(-\gamma\|t_i -t_j\|^2)$ with the bandwidth $\gamma$, and $d^2:\mathbb{R}^{3}\times\mathbb{R}^{3}\rightarrow\mathbb{R}$ is the geodesic distance between $\mathbf{s}^x(t_i)$ and $\mathbf{s}^y(t_j)$ which reads $\arccos(\mathbf{s}^x(t_i)^\top\mathbf{s}^y(t_j))$. 
	The overall cost is reweighted by the weights $w(i)$ to highlight the cost where the signal has high intensity and vice versa. 
	This is necessary because the normalized Stokes parameters are more susceptible to the noise in the regions where the intensity $S^0_x(t)$ is small w.r.t. the noise. 
	Due to the high penalty evaluated in those regions, the solution is unnecessarily penalized in the regions where the signal is less present. 
	In order to correct this, we propose to use the weights in Eq.~\eqref{eq:secprob}.
	Finally, we obtain the penalty term $\ell_k(\mathbf{x}|\mathbf{y})$ which is minimized when the normalized Stokes parameters of the solution are both locally smooth in time and close to that of the measurements.  
	Here both the fidelity (likelihood) and the regularization (prior) terms are combined into a single penalty term and $\gamma$ and $W$ determine the bandwidth of the locality. 
	Higher values yield more global smoothing in time. 
	Given this new penalty term, we propose to modify the problem definition of the previous section by replacing the Stokes parameters with the normalized ones: 
	\begin{dmath}
		\mathbf{x}^{\star}_{ker} \small= \argmin_{\mathbf{x}\in\mathbb{R}^{2\times N}} f_1(\mathbf{x},\mathbf{y})+ \beta_1\sum_{k=1}^{N}\ell(y(t_k)|x(t_k)) + \lambda_s(\mathbf{S}_{0}^x)^\top\mathsf{L}\mathbf{S}_{0}^x + \alpha \ell_k(\mathbf{x}|\mathbf{y}),
		\label{eq:secprob}
	\end{dmath}	 
	where $\alpha$ is the scalar to control the effect of the kernel regres-
	sion penalty.
	The resulting cost function is also differentiable which can be tackled in a similar manner with the previous case. 
	The solution $\mathbf{x}^{\star}_{ker}$ is locally smooth in time both in individual components and the intensity and normalized Stokes parameters. 
	However, it is not the same as the solution of Eq.~\eqref{eq:firstprob} as expected. 
	The main advantage of this formulation is that it allows us to regularize the normalized Stokes parameters on the sphere over time. 
	In addition, the parameters $\gamma$ and $W$, and the kernel function $K$ allow us to directly determine the locality to which we would like to apply the smoothness. 
	In the next section, we illustrate both frameworks and compare their weak and strong points in the numerical simulations. 
	\begin{figure}
	\centering
	\includegraphics[width=1.0\linewidth]{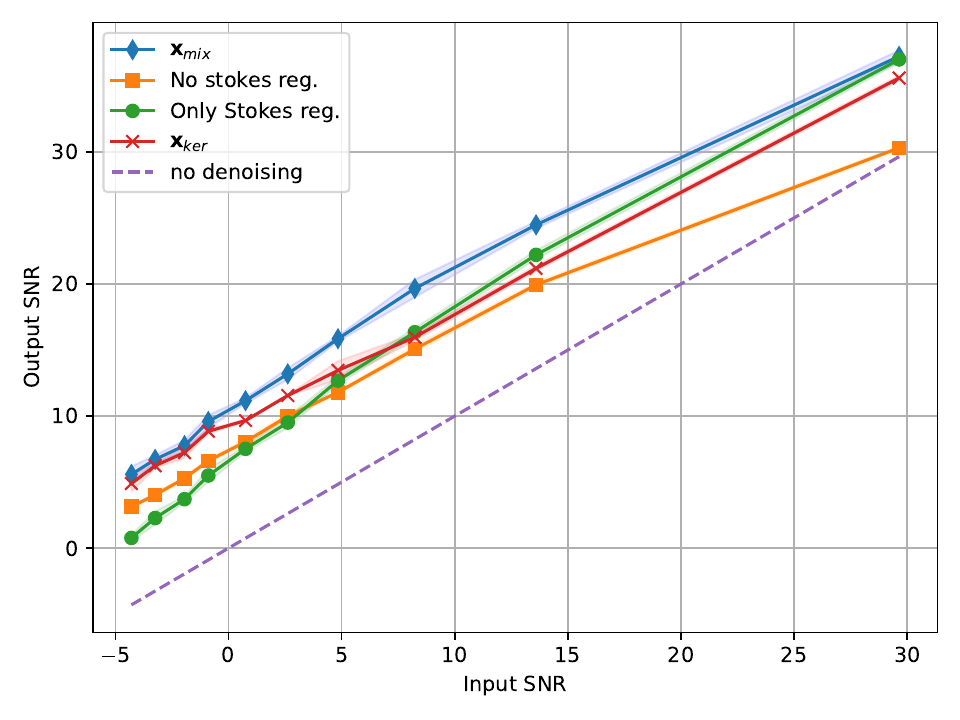}
	\caption{The performance of the solutions to the first (in blue, orange and green) and second formulations (in red) in Eq.~\eqref{eq:firstprob} and Eq.~\eqref{eq:secprob}. The different parameter settings are used for the first formulations; $(\lambda_1>0,\lambda_s>0,\beta_1>0,\beta_2>0)$ in blue curve, $(\lambda_1>0,\lambda_s=0,\beta_1=0,\beta_2=0)$ in orange curve and $(\lambda_1=0,\lambda_s>0,\beta_1>0,\beta_2>0)$ in green curve. }
	\label{fig:result1}
\end{figure}

	\section{Numerical Simulations}
	\label{sec:experiments}
	In this section, we show the performance of the proposed frameworks on a synthetic example. 
	At first, we generate an AM-FM polarized bivariate signal. The time-varying parameters are chosen as follows; $a(t)$ is the Hanning window with $N=1024$ samples, $\phi(t)= 2\pi f_0 t$, $\theta(t)= \pi/16-t$, $\chi(t)= \pi/4 - 2t$, 	
	where $f_0=15.90$ and $t\in[0,2\pi/4]$. 
	After generating the true signal $\mathbf{x}$ for $N=1024$ consecutive timestamps, we generate the noisy measurements by adding artificial Gaussian noise with the standard deviation $\sigma\in\{0.01,\dots,0.50\}$. 
	Finally, we run the ADAM gradient descent algorithm~\cite{kingma2014adam} to solve the problem in Eq.~\eqref{eq:firstprob} and~\eqref{eq:secprob} for different realizations of noisy signals. 
	To see the effect of the Stokes-based regularization, we solve the problem in~\eqref{eq:firstprob} for three cases; \textbf{i/} individual component regularization~\ie~$(\lambda_1>0,\lambda_s=0,\beta_1=0,\beta_2=0)$, \textbf{ii/} only Stokes regularization~\ie~$(\lambda_1=0,\lambda_s>0,\beta_1>0,\beta_2>0)$, \textbf{iii/} both Stokes and individual component regularization~\ie~$(\lambda>0,\lambda_s>0,\beta_1>0,\beta_2>0)$. 
	While we plug the true $\sigma$ value in our frameworks, the other hyperparameters are tuned via grid search for each case and the input vs. output signal-to-noise ratio is plotted in Fig.~\ref{fig:result1}. 
	The locality parameters for the kernel regression loss are chosen as $\gamma=0.2$ and $W=32$.  
	In this numerical example, one can easily see that the best performances are achieved when one has regularization on both the signal and Stokes domain. 
	\begin{figure}
		\includegraphics[width=\linewidth]{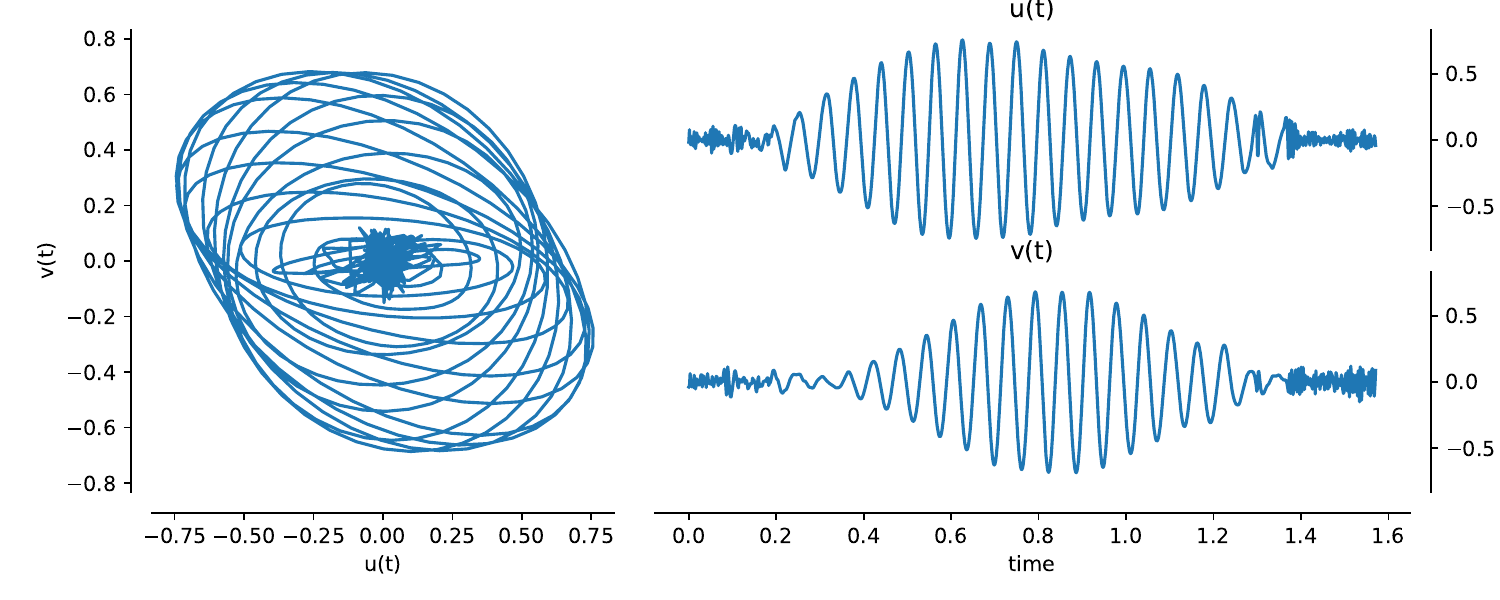}
		\includegraphics[width=\linewidth]{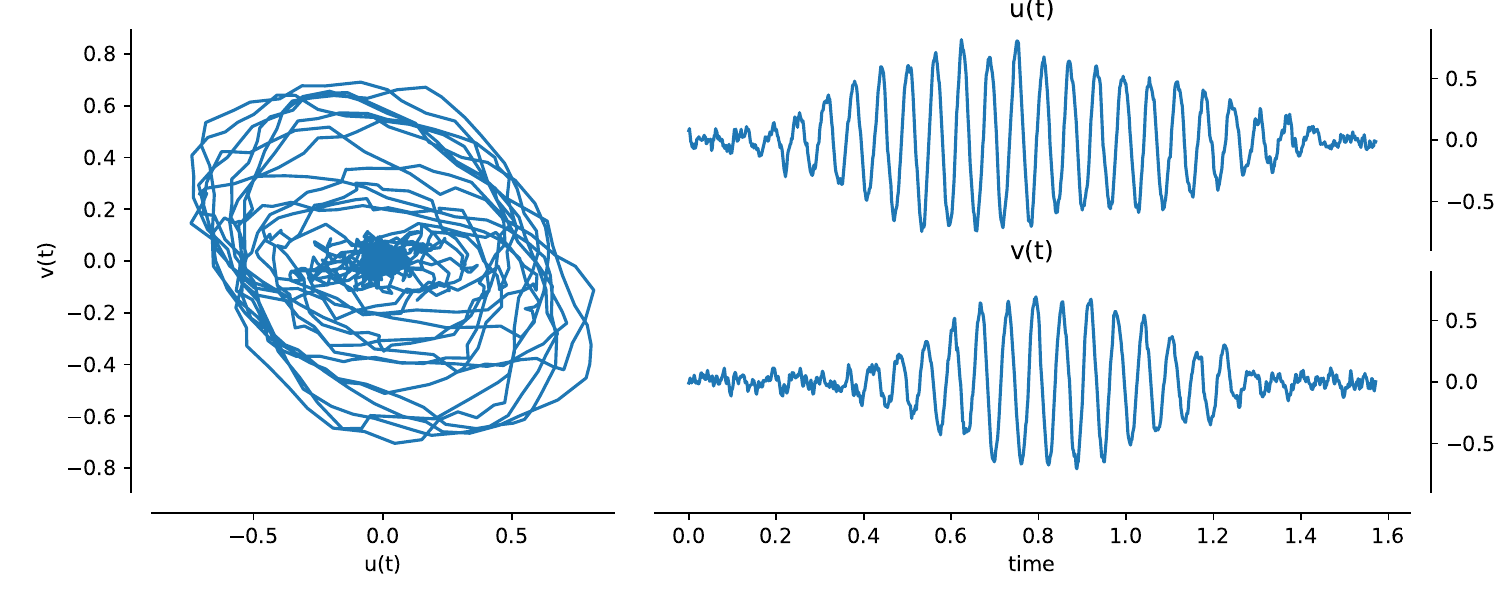}
		\includegraphics[width=\linewidth]{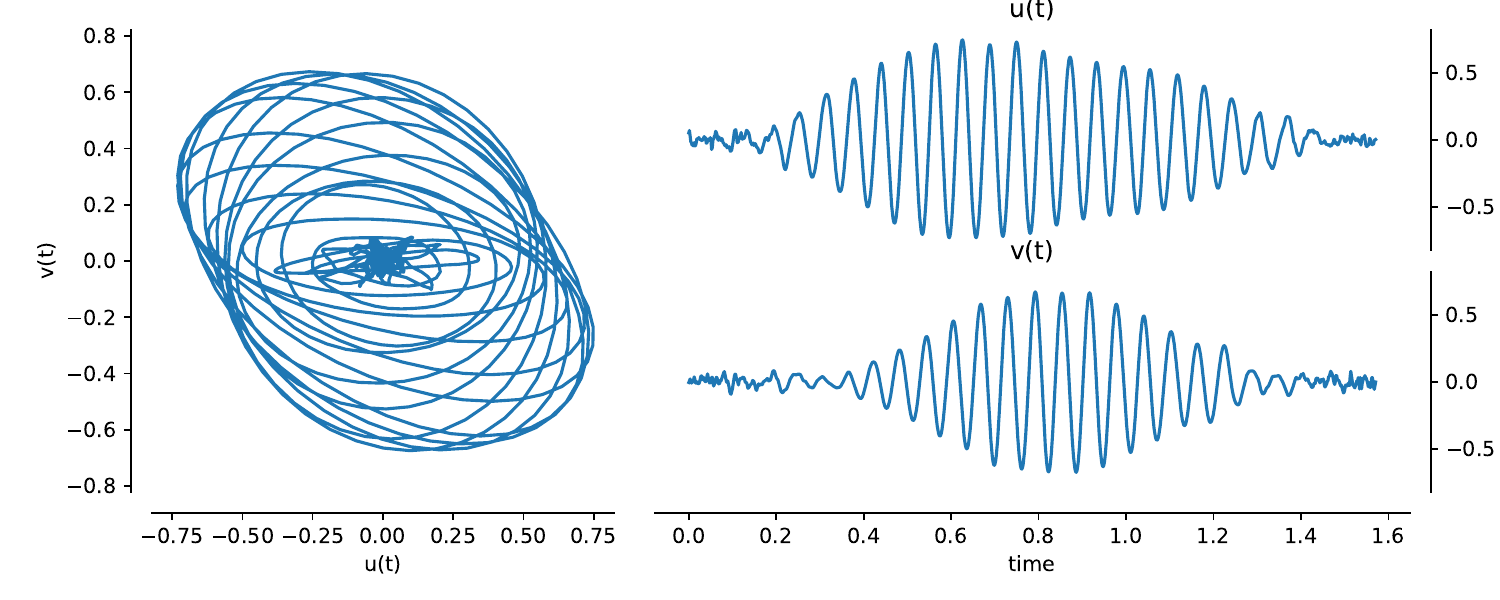}
		\caption{The solution to the problem in Eq.~\eqref{eq:firstprob} with $(\lambda_1=0,\lambda_s>0,\beta_1>0,\beta_2>0)$ at top,$(\lambda_1>0,\lambda_s=0,\beta_1=0,\beta_2=0)$ at center and $(\lambda_1>0,\lambda_s>0,\beta_1>0,\beta_2>0)$ at bottom. The SNRs are $18.03$, $16.44$ and $20.62$ from top to bottom. The true and noisy signal is given in Fig.~\ref{fig:bivariate-example}.}
		\label{fig:result2}
	\end{figure}
	A closer look in Fig.~\ref{fig:result2} shows that smoothing the components individually cannot directly regulate their interrelation as the smooth signals do not necessarily follow the same polarization properties as the true signal. 
	Similarly, the signal regularized in the Stokes level might yield a signal that is highly fluctuating as the Stokes parameters are energetic quantities that do not carry any information related to the phase parameter of the signal~\cite{flamant2018approche}. 
	As a result, the regularization in both domains is highly necessary to recover the true signal. 
	\begin{figure}
		\centering
		\includegraphics[width=0.9\linewidth,trim={1.5cm 0cm 1.5cm 0cm},clip]{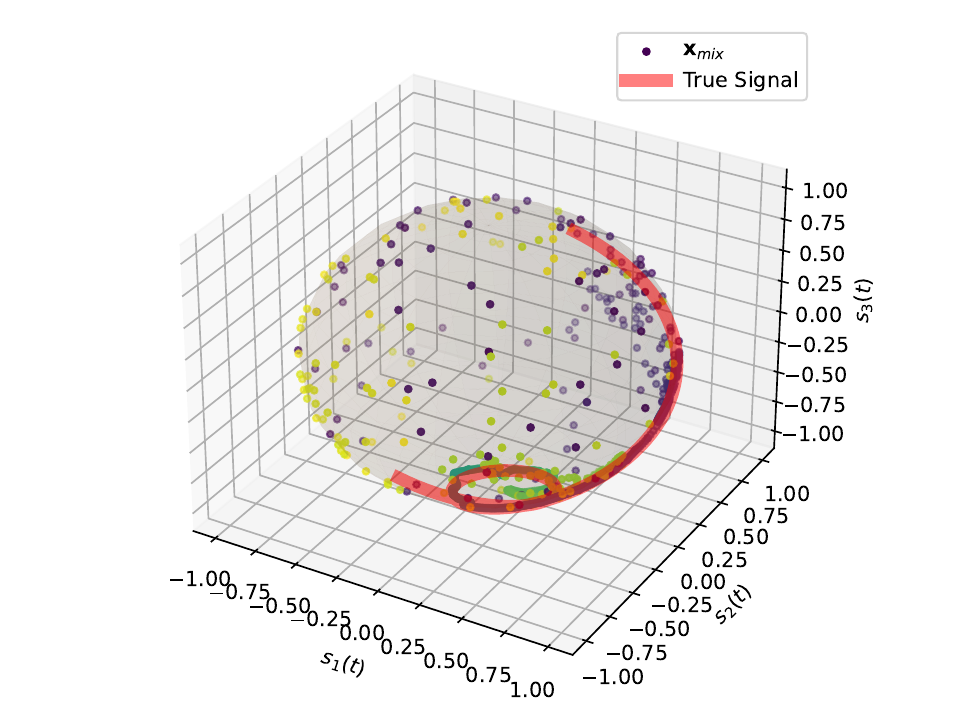}
		\includegraphics[width=0.9\linewidth,trim={1.5cm 0cm 1.5cm 0cm},clip]{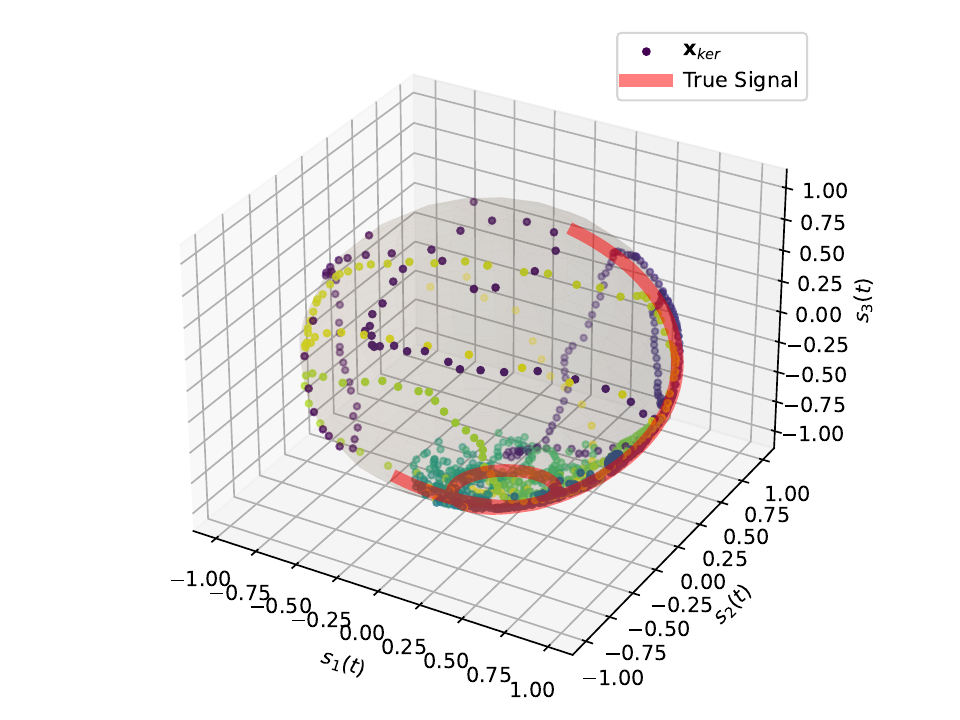}
		\caption{The normalized Stokes parameters $\mathbf{s}^x$ of $\mathbf{x}_{mix}$ (top), $\mathbf{x}_{ker}$ (bottom)
			and the true signal (in red in both figures) is plotted on the unit sphere.
			Each point corresponds to a vector $\mathbf{s}^x(t_i)$, $i = 1, \hdots, N$ , of $\mathbf{x}_{mix}$ and
			$\mathbf{x}_{ker}$ and the plotted parameters are colored to indicate their timestamps. The
			timestamps increase from blue to yellow point.}
		\label{fig:on_sphere}
	\end{figure}
	The comparison between $\mathbf{x}_{mix}$ and $\mathbf{x}_{ker}$ shows that $\mathbf{x}_{mix}$ usually performs slightly better than $\mathbf{x}_{ker}$ while both successfully leverages the signal and Stokes domain information. The underlying reason can be investigated by looking at the normalized Stokes parameters of these two solutions in Fig.~\ref{fig:on_sphere}.  
	The normalized Stokes parameters of $\mathbf{x}_{mix}$ are much closer to that of the true signal where the signal has high intensity and is completely noisy at the beginning and end of the time axis where the signal has low intensity. 
	On the other hand, the difference of the normalized Stokes parameters of $\mathbf{x}_{ker}$ from the true parameters is less relevant to the intensity. 
	This is expected due to the reweighting scheme in Eq.~\eqref{eq:secprob}.  
	However, the denoising performance of $\mathbf{x}_{ker}$ is still inferior probably due to the intrinsic noise susceptibility of the normalized Stokes parameters that is caused by the normalization.  Further investigation can be done by analyzing the theoretical or empirical distributions of the normalized parameters by going beyond the scope of this work. 
	\section{Conclusion}
	\label{sec:conc}
	We propose two formulations for solving an inverse problem for bivariate signals called denoising. These formulations allow us to leverage both intra and inter-component statistics and they are based on theoretically solid statistical frameworks, namely Bayesian inference and kernel regression.
	Both in these formulations, the inter-component statistics are represented via the Stokes parameters which are closely related to the polarization properties of the bivariate signals. Therefore, they are particularly useful and more interpretable in denoising problems under the assumption that the true signal is polarized while the noise is not. The numerical simulations on a synthetic signal strongly validate this hypothesis. 
	
	% This work opens many new avenues for the future. 
	% For example, a direction of broad interest is to extend these frameworks to the multi-frequency component signals. This requires to to move our analysis to the time-frequency Stokes representations. 
	% Another direction can be taken in the algorithmic side by developing more suitable optimization algorithms for the formulations at hand, possibly using quaternion optimization tools~\cite{flamant2021general}.    

	\section*{Acknowledgment}
	The authors acknowledge support from the ANR project ``Chaire IA Sherlock'' ANR-20-CHIA-0031-01, the {\em programme d'investissements d'avenir} ANR-16-IDEX-0004 ULNE and Région Hauts-de-France. 
	This research has been supported in whole or in part, by the Agence Nationale de la Recherche (ANR) under project ANR-21-CE48-0013-03. In the interest of open access publication, the author/rights holder applies a CC- BY open access license to any article/manuscript accepted for publication (AAM) resulting from this submission.
	\bibliographystyle{abbrv}
	\bibliography{refs}
\end{document}